# Growth and Superconductivity of FeSe$_x$ Crystals

U. Patel, J. Hua, S. H. Yu, S. Avci, and Z. L. Xiao*

Department of Physics, Northern Illinois University, DeKalb, Illinois 60115

H. Claus, J. Schlueter, V. V. Vlasko-Vlasov, U. Welp, and W. K. Kwok

Materials Science Division, Argonne National Laboratory, Argonne, Illinois 60439

Iron selenide (FeSe$_x$) crystals with lateral dimensions up to millimeters were grown via a vapor self-transport method. The crystals consist of the dominant $\alpha$ - phase with trace amounts of $\beta$ - phase as identified by powder x-ray diffraction. With four-probe resistance measurements we obtained a zero-resistance critical temperature of 7.5 K and a superconducting onset transition temperature of up to 11.8 K in zero magnetic field as well as an anisotropy of 1.5 ± 0.1 for the critical field. Magnetization measurements on individual crystals reveal the co-existence of superconductivity and ferromagnetism.

Studies on iron selenides, FeSe$_x$, traces back more than 70 years, driven by the pursuit of semiconducting and magnetic properties in transition metal binary chalcogenides[1-3]. In recent years, research in FeSe$_x$ has intensified due to their potential applications as spin injection materials in spintronics[4-7]. At room temperature FeSe$_x$ is stable in the PbO type structure ($\alpha$ - phase) with $x$ up to 8/7 and in the NiAs type structure ($\beta$ - phase) with $x \approx$ 8/7~4/3 [Ref.1]. Magnetization measurements indicate that in the $\beta$ - phase, FeSe$_x$ is ferromagnetic/ferrimagnetic at or above liquid nitrogen temperatures[2,7] while in the $\alpha$ - phase, both the Fe deficient[5-7] and Se deficient[7] material is ferromagnetic near room temperature.

Stimulated by the recent discovery of high temperature superconductivity in iron-based layered quaternary oxypnictides RFeAsO$_{1-x}$F$_x$ (R = La, Nd, Sm etc.)[8-10], other iron-based planar compounds have been revisited to search for superconductivity. $\alpha$ - FeSe$_x$ has the planar crystal sublattice consisting of edge sharing FeSe$_4$ tetrahedra, the same as the FeAs$_4$ tetrahedra layers found in oxypnictides[11]. Thus discovery of superconductivity in FeSe$_x$ will be valuable in understanding the superconducting mechanism of Fe-based superconductors. Furthermore, FeSe$_x$ is much easier to handle and fabricate since it is a binary system and selenium is much less toxic than arsenic. By exploring the effect of composition, Hsu et al. indeed observed superconductivity with a zero resistance superconducting critical temperature ($T_{c0}$) of ~ 8 K and an onset critical temperature ($T_c^{on}$) of ~ 12 K in Se-deficient FeSe$_{0.82}$ and FeSe$_{0.88}$ [Ref.11]. The superconductivity of FeSe$_x$ can be further enhanced by partially replacing Se with Te [Ref.12] which led to a $T_{c0}$ of up to 14 K. High pressure studies at 1.48GPa [13] have yielded $T_{c0}$ and $T_c^{on}$ of 13.5 K and 27 K, respectively. Investigations on the superconductivity in FeSe$_x$ [Refs. 14-16] can shed light on the role of FeAs layers and magnetic interactions related with the occurrence of superconductivity[14] in the recently discovered superconducting ferrous-oxypnictides.

Currently reported experiments were carried out with powder samples synthesized using a procedure reported by Hsu et al., which involves sintering high purity elemental Fe and Se powders in an evacuated quartz tube. Under this synthesis, impurity phases such as iron, iron oxide, iron silicide[11], $\beta$ - FeSe$_x$ [Refs. 11,14] were found and can even dominate[14] in powder samples. Flux growth of FeSe$_x$ single crystals following sintering of FeSe$_x$ powder with NaCl/KCl flux[17] has resulted in FeSe$_x$ crystals with lateral dimensions of a few hundreds micrometers. The flux-grown FeSe$_x$ crystals show zero resistance at $T_{c0}$ = 3.4 K with a superconductivity onset at $T_c^{on}$ = 11.9 K. The superconducting transition thus much broader and the $T_{c0}$ is much lower than those reported in powder samples[11,13,14].

This Letter reports the growth of FeSe$_x$ crystals using a vapor self-transport approach which avoids the potential flux contamination and water induced degradation of sample quality encountered with the flux method. $\alpha$ - FeSe$_x$ crystals with lateral dimensions up to a few millimeters were obtained and their superconducting critical temperatures and transition widths were found to be remarkably consistent with that reported in powder samples. Resistive measurements indicate an anisotropy of the critical field. We also observed a ferromagnetic background in the superconducting state.

In the vapor transport growth of crystals, the diameter and temperature profile of the ampoule, the nucleation time, the transport agent, and the pressure among other variables could affect the quality and size of the crystals. Thus significant efforts are required to optimize the growth conditions. We tried a well-established iodine vapor transport (IVT) approach for growing large NbSe$_2$ crystals[18]. No crystals of FeSe$_x$ were found after one-month growth duration. However, a repetition of the same procedure without iodine yielded crystals of FeSe$_x$, indicating that iodine likely reacts with Fe and hinders the formation of FeSe$_x$ crystals.

High purity iron (> 99.9%) and selenium (> 99.99%) powders were combined with desired composition ratios. The mixture was ground and sealed in an evacuated 30 cm long, 14 mm (ID) quartz ampoule after purging repeatedly with high purity Ar gas to ensure an oxygen free environment. Ampoules were then heated in a three-zone furnace with the following procedure[18]: i) the growth region was cleaned with a temperature setting of 700 °C, 900 °C and 900 °C for 5 days (with the Fe & Se mixture located at one end of the furnace at a temperature of 700 °C); ii) the crystal nucleation was initiated with a zero temperature gradient with all zones held at 700 °C for 5 hours; iii) the crystals were left growing for 30 days with a three zone temperature setting of 825 °C, 700 °C, and 825 °C respectively. The cool-down procedure is to decrease the temperature slowly (at a rate of 3 °C/min) to room temperature after keeping the samples at 400 °C for 10 hours. Powder x-ray diffraction (XRD) analysis with a monochromatic CuK$_a$ ($\lambda$ = 1.540598 Å) radiation source was used to determine the phase purity of the crystals. The pictures of the crystals were taken with an optic microscope. Resistive and magnetization measurements were carried out in a Quantum Design PPMS-9 and MPMS-7, respectively.

Using a flux-growth approach, Zhang et al. obtained crystals with tetragonal and hexagonal shapes consisting of $\alpha$ - and $\beta$ - phase, respectively[17]. With our synthesis approach, we obtained crystals with hexagonal and trapezoidal shapes in the same quartz tube. The four insets to Fig.1 present images of typical as-grown crystals with nominal lateral dimensions ranging from a few hundred micrometers to 1-2 mm and thickness of tens of micrometers. As indicated in the main panel of Fig.1, XRD patterns obtained from a few pieces of crystals with the same shape indicate that the $\alpha$ - phase dominates in crystals of both shapes, although trace amounts of $\beta$ - phase can also be detected. The derived lattice constants of the $\alpha$ - phase are $a = b = 3.7794$ Å, $c = 5.5209$ Å and $a = b = 3.7748$ Å, $c = 5.5243$ Å for the crystals with trapezoidal and hexagonal shapes, respectively. Energy dispersive x-ray spectroscopy analyses (Hitachi S-4700) indicate that crystals of both shapes have the same composition (within the experimental error of 2-3%).

We conducted four-probe resistance measurements on individual trapezoidal and hexagonal shaped crystals. Since $\alpha$ - FeSe$_x$ is a layer structured with a planar crystal sublattice, its superconducting properties can be anisotropic. As revealed by the data presented in Fig.2(a), the angle dependence of the resistance ($R$ - $\theta$) obtained at a fixed temperature ($T = 8$ K) under various magnetic fields for a trapezoidal shaped crystal indeed shows minima and maxima. We obtained the maximum anisotropy value from the temperature dependence of the resistance ($R$ - $T$) curves at $\theta_M = 90°$ and $\theta_m = 155°$ where the resistance is at a maximum and a minimum, respectively. The main panel of Fig.2(b) presents $R$ - $T$ curves at $\theta_m = 155°$. The zero-field $R$ - $T$ curve showing a $T_{c0}$ of 7.5 K and a broad transition remarkably resembles that reported by Hsu et al. for FeSe$_x$ powder samples. This indicates that the broad superconducting transition in their powder samples could be an intrinsic property of individual crystalline grains instead of the distribution of critical temperature among different grains. Using the same criterion of 50%$R_N$ as what was used in powder samples to define the critical temperature $T_c$, we obtained the phase diagram at the $\theta_m = 155°$ magnetic field direction as shown in the lower-right inset of Fig.2(b) as squares which gives a zero-field $T_c$ of 9.44 K. Our crystal has an estimated zero temperature critical field $H_{c2}(0)$ of 37.5 T which is more than twice that (16.3 T) of powder samples. One possible reason could be that in the powder samples the magnetic field was not in the lowest dissipation direction or the highest dissipation grains dominated the transport properties. Indeed, the circles in the same inset depicting the phase diagram shows that in a magnetic field direction of $\theta_M = 90°$ where the dissipation is at maximum, the estimated $H_{c2}(0)$ is 24.9 T which is much closer to that of the powder samples. Similarly, the phase diagram presented in the upper-left inset of Fig.2(b) for a hexagonal shaped crystal depicts a slightly lower $T_c$ of 9.16 K in the absence of magnetic field and an $H_{c2}(0)$ of 36.5 T and 24.5 T for the minimal and maximal dissipation directions, respectively. The ratios of the critical fields obtained with these two magnetic field angles lead to an anisotropy of 1.506 and 1.490 for the trapezoidal and hexagonal shaped crystals, respectively. That is, the anisotropy of FeSe$_x$ determined from the field directions where the dissipation is at a minimum and at a maximum is 1.50 ± 0.01. However, $\theta_M$ and $\theta_m$ change from sample to sample. This indicates that the anisotropy of the critical field can be extrinsic, for example, due to surface superconductivity.

Neutron scattering experiments demonstrate that the superconductivity in iron-based compound may be associated with intrinsic magnetism[12]. First principal calculations[16] also indicate the existence of magnetism in FeSe$_x$ driven by Se vacancy. In fact, a ferromagnetic magnetization loop was observed in Se deficient $\alpha$ - FeSe$_x$ thin films at room temperature, though both experiments and calculations indicated that the stoichiometric phase is nonmagnetic[7]. Thus, magnetization characterizations on FeSe$_x$ crystals will be important not only to identify its superconductivity related diamagnetism but also likely to provide information on the superconducting mechanism in iron-based superconductors. In fact, a large positive background in the susceptibility was observed in the normal state of both FeSe$_x$ powders and flux-grown crystals[11, 17] and attributed to the existence of Fe impurity[11].

The upper-left inset of Fig.3 presents the magnetization versus temperature (M - T) curves obtained in a magnetic field of 30 G for a trapezoidal shaped crystal. The magnetization starts to deviate from a positive constant background at a temperature around 8 K that is consistent with the $T_{c0}$ obtained from resistance measurements. The magnetization versus magnetic field ($M - H$) loop obtained for a trapezoidal shaped crystal at 12 K (Fig. 3 main panel) shows that the background is ferromagnetic. In contrast to the powder samples[11], the ferromagnetic background can be seen in the superconducting state, as demonstrated by the $M - H$ loops for crystals of both shapes at 4 K. Since XRD analyses indicate that our crystals have no iron, iron oxide or iron silicide impurities as reported in the powder samples, we surmise that the ferromagnetic background could be a result of either intrinsic magnetism of $\alpha$ - FeSe$_x$ crystals due to Se vacancy[16] or trace amounts of $\beta$ - phase which is ferri-/ferromagnetic[7]. Further work on magnetization of pure $\alpha$ and $\beta$ - phases at low temperatures is needed to clarify the situation.

In summary, we grew $\alpha$ - phase dominant FeSe$_x$ crystals with lateral dimensions up to millimeters. Both resistance and magnetization measurements on individual crystals reveal superconductivity behavior comparable to that reported in powder samples. A magnetic field anisotropy of ~1.50 is determined between the maximal and minimal dissipation directions. Magnetization measurements also reveal the co-existence of superconductivity and ferromagnetism.

This work was supported by the US Department of Energy Grant No. DE-FG02-06ER46334 and Contract No. DE-AC02-06CH11357. The compositional and morphological analyses were taken at Argonne's Electron Microscopy Center (EMC) and the Center for Nanoscale Materials (CNM), respectively.

*Corresponding author, zxiao@niu.edu or xiao@anl.gov

**Figure captions**

**FIG.1.** (color online) XRD patterns of FeSe$_x$ crystals. Top and bottom curves are obtained from a few pieces of hexagonal and trapezoidal shaped crystals grown from a starting powder mixture with Fe/Se ratio of 1:0.82, respectively. α - phase dominates in both types of crystals while trace amounts of β - phase (marked with asterisks) can be identified. The insets show typical optical images of the corresponding crystals. The scale bar is 200 μm.

**FIG.2.** (color online) (a) $R$ - $\theta$ curves at $T$ = 8 K for various magnetic fields (1T - 8T with a interval of 1T) where $\theta$ = 0° is defined as the magnetic field direction perpendicular to the crystal plate (see the inset for the definition). (b) $R$ - $T$ curves at various magnetic fields applied at an angle $\theta_m$ = 155° where the dissipation is at a minimum. The phase diagrams of trapezoidal and hexagonal crystals obtained at field directions of maximal (circles) and minimal (squares) dissipation are presented in the lower-right and upper-left insets, respectively. The magnetic field rotates in a plane perpendicular to the current.

**FIG.3.** (color online) $M$ - $H$ loops of a trapezoidal shaped crystal at $T$ = 4 K and 12 K. The upper-left inset give the $M$ - $T$ curve for a trapezoidal shaped crystal at $H$ = 30 G. Lower-right inset shows the $M$ - $H$ loop of a hexagonal shaped crystal at $T$ = 4 K.

**FIG.1**

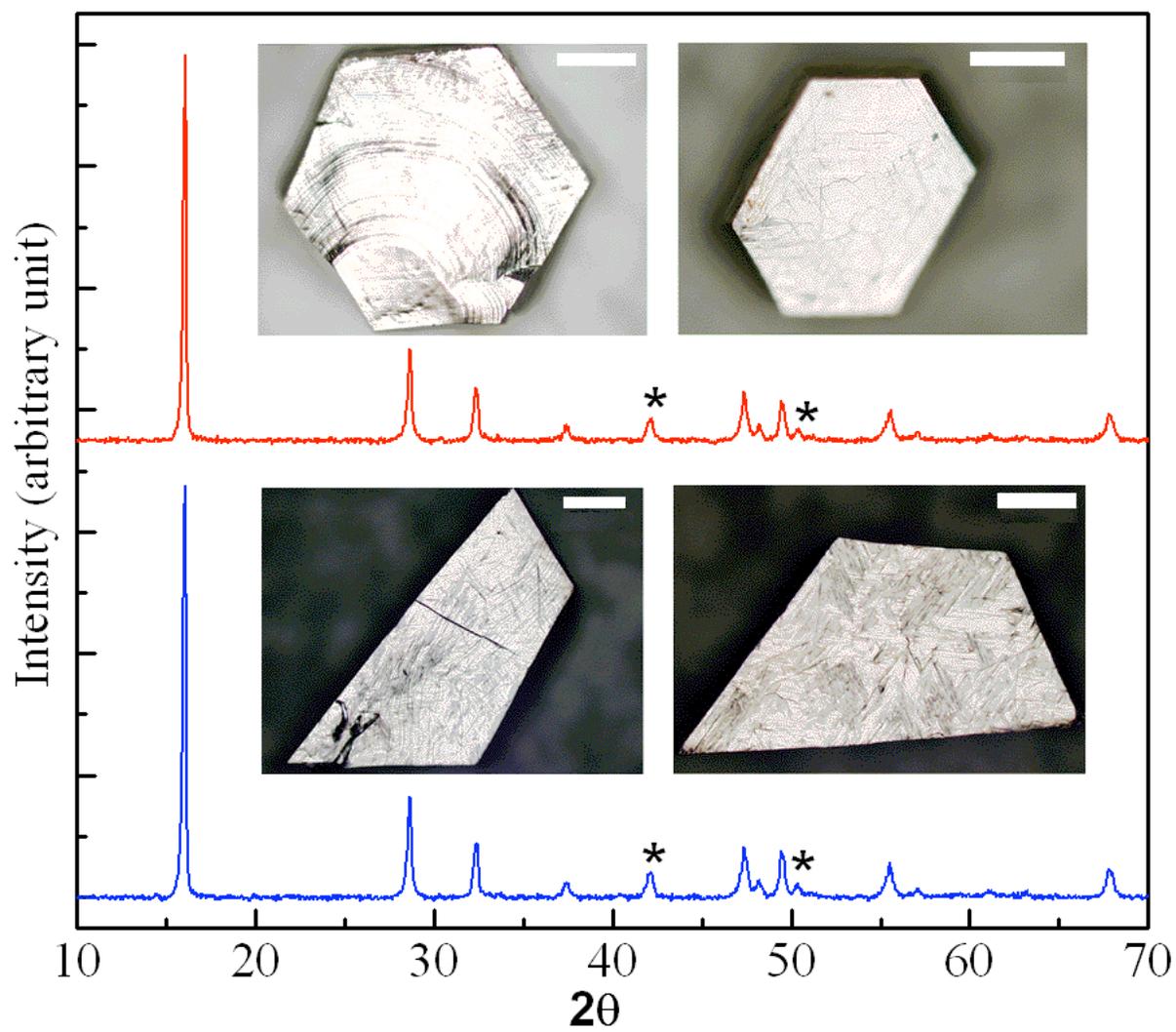

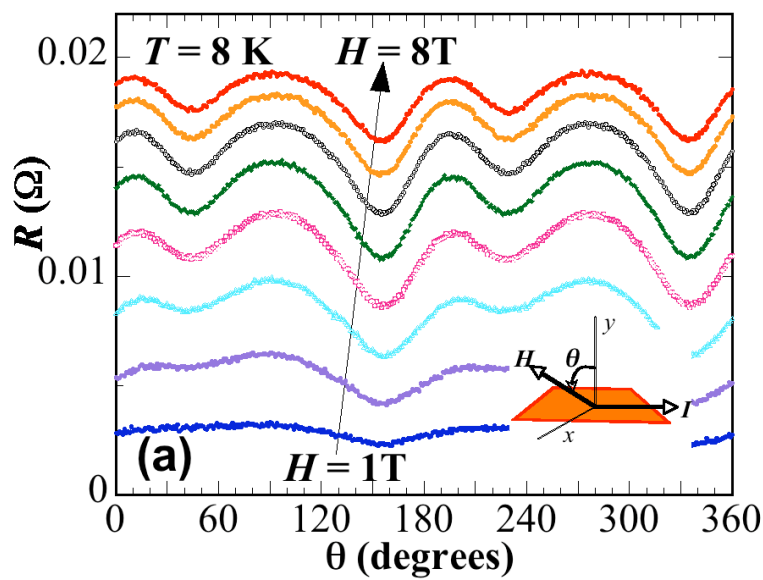

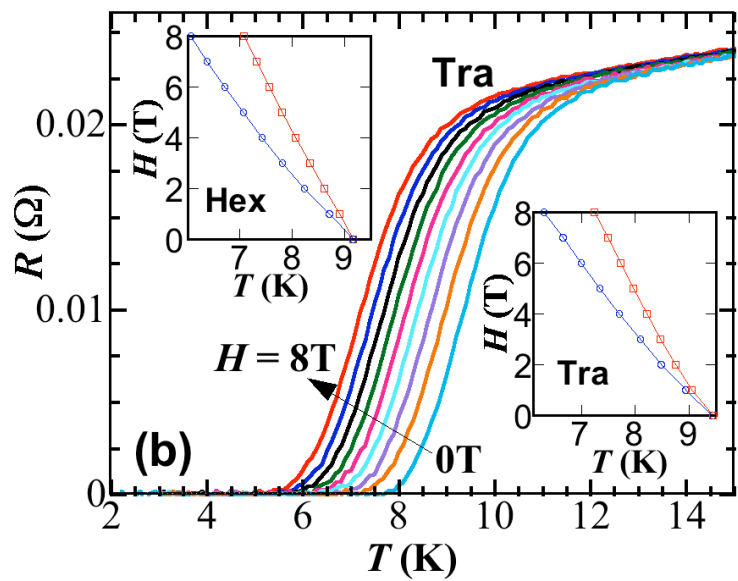

**FIG.3**

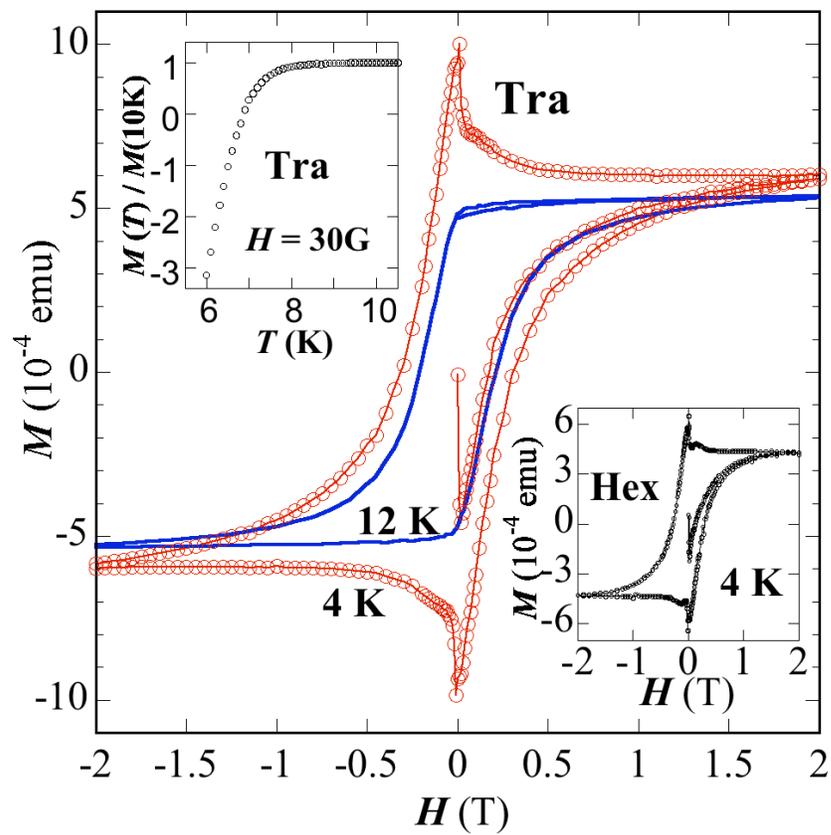